\documentclass[aps,pra,reprint,superscriptaddress,showpacs,longbibliography]{revtex4-1}
\usepackage{braket}
\usepackage{amsmath}
\usepackage{amsfonts}
\usepackage{amssymb}
\usepackage{graphicx}
\usepackage{color}
\usepackage{dcolumn}
\usepackage{bm}
\usepackage{standalone}

\begin{document}



\title{Quantum particle in a split box: Excitations to the ground state}



\author{Vegard~S{\o}rdal}
\affiliation{Department of Physics, University of Oslo, 0316 Oslo, Norway}
%

\author{Joakim~Bergli}
\affiliation{Department of Physics, University of Oslo, 0316 Oslo, Norway} 


\date{\today}

\begin{abstract}
We discuss two different approaches for splitting the wavefunction of a single-particle-box (SPB) into two equal parts. Adiabatic insertion of a barrier in the center of a SPB in order to make two compartments which each have probability 1/2 to find the particle in it is one of the key steps for a Szilard engine. However, any asymmetry between the volume of the compartments due to an off-center insertion of the barrier results in a particle that is fully localized in the larger compartment, in the adiabatic limit. We show that rather than exactly splitting the eigenfunctions in half by a symmetric barrier, one can use a non-adiabatic insertion of an asymmetric barrier to induce excitations to the first excited state of the full box. As the barrier height goes to infinity the excited state of the full box becomes the ground state of one of the new boxes. Thus, we can achieve close to exact splitting of the probability between the two compartments using the more realistic non-adiabatic, not perfectly centered barrier, rather than the idealized adiabatic and central barrier normally assumed.
\end{abstract}

\maketitle

\section{Introduction}

The Szilard engine is a simple conceptual model of a information processing system. The classical model is a single particle in a box, coupled to a thermal bath. By inserting a movable barrier in the center of the box the probability to find the particle in either compartment becomes 1/2. If we now perform a measurement to find out which compartment the particle is in, we generate one bit of Shannon information which is stored in some memory. Since the box is coupled to a thermal bath, we can extract work by allowing the compartment we find the particle in to expand and fill the whole box. The maximum work extracted in this way is $ k_BT\log 2 $, and this is achieved by reversible expansion. To complete the cycle the memory is deleted, which has a minimum energy cost of $ k_BT \log 2 $, according to Landauers principle. Therefore, if we preform reversible operations, the full cycle of measurement, work extraction and information deletion generates no entropy. The quantum mechanical version of the Szilard engine is similar, only now we are splitting the wave function of the particle. The quantum measurement and, assuming the memory is classical, deletion is similar to the classical case, but there are subtle differences when it comes to the insertion, expansion and removal of the barrier \cite{PhysRevLett.106.070401}.\\

The adiabatic theorem in quantum mechanics tells us that a system remains in its instantaneous eigenstate as long as it has a gapped energy spectrum and the perturbation acting on it is slow enough to prevent transition between the eigenstates. Based on this, it has been remarked in \cite{AmJourPhys} that if the particle is in the ground state and the barrier is inserted off-center, such that one compartment is larger than the other, the particle will always be localized in the larger compartment. This is because the energy spectrum is proportional to $ L^{-2} $, where L is the length of the compartment. The result is independent of how small the asymmetry between the compartments are; any finite difference between the compartment sizes will give the same result.\\

With modern technology we can now experimentally realize what was before only a thought-experiment. The creation of Szilard engines in a range of physical systems have been reported the last decade: atoms~\cite{raizen2008,thorn2008,Raizen2009}, colloidal 
particles~\cite{toyabeNPhys2010,berut2012}, molecules~\cite{serreli2007},
electrons~\cite{Koski2014,Koski2015,chida2015}, and
photons~\cite{vidrighin2016}. In experiments the barrier is not inserted adiabatically, nor exactly in the center, and one can ask the question how the result of the previous paragraph changes when it is inserted at a finite rate. The result is that the particle is not fully localized in either compartment.\\

Although a fast rate of insertion can make the probability to find the particle in the smaller compartment non-zero, the  downside is that a fast rate results in excitations to higher energy levels. The Szilard engine measurement procedure traditionally only determines which side of the box the particle is found, not is exact eigenstate. Excitation of high energy levels introduces additional entropy that is not accounted for in the which-side measurement. Therefore in the full Szilard engine operating cycle of barrier insertion, measurement, extraction of work and deletion of memory, has decreased efficiency. \\

A good thought-experiment is never set in some complicated system with many degrees of freedom. Rather, it is a surprising result or counterintuitive implication obtained from the study of a simplified model of reality. One might ask why further study of a thought-experiment that has already been experimentally realized is necessary. In our opinion there are two main reasons: \\
The first reason is that studying all the aspects of this conceptual model helps us to understand the key physical effects that gave rise to the thought-experiment in the first place, and guides us in how to think about their order of importance. \\
The second reason is that even though thought-experiments can guide our understanding regardless of whether it is possible to experimentally perform them it is also important to investigate whether they present practical possibilities. This point is especially relevant for the Szilard engine, which could be used as a model for information processing devices in future technology.\\

\begin{figure}
\centering
\includegraphics[width=1\linewidth]{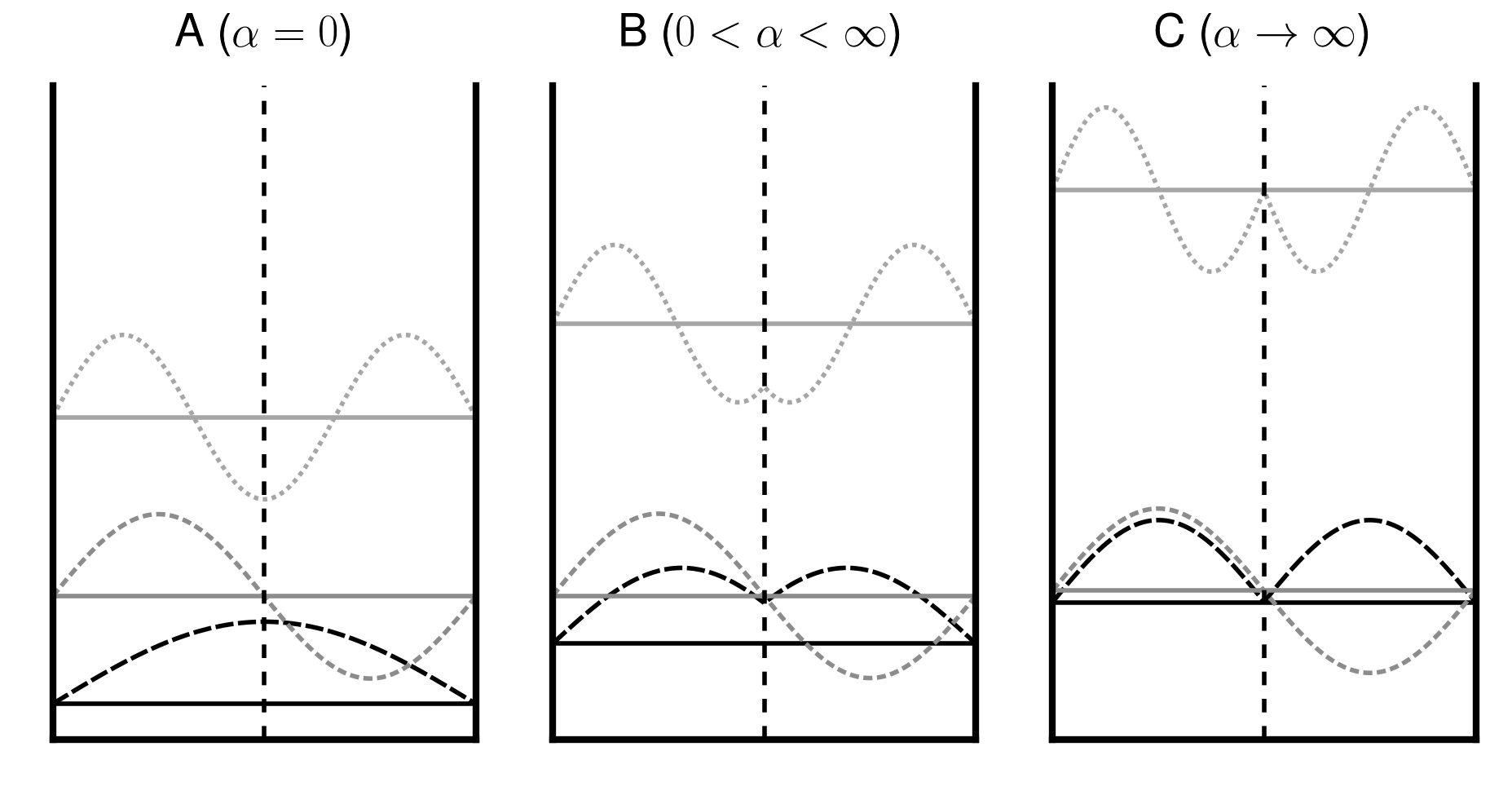}
\caption{Schematic of the three first eigenfunction and energies for a symmetric box for three different values of the barrier height $ \alpha(t) $. In A) we show the initial state of the system, before the barrier is inserted. B) is at an intermediate time before $ \alpha(t)\to \infty $. We see that when the barrier is inserted at the center of the box it hits the nodes of the antisymmetric eigenfuntions, and therefore there are no excitations to this state (see Eq.~(\ref{coeff})). C) shows the limit when $ \alpha(t)\to\infty $. The total wavefunction is symmetric about the barrier, and the probability to find the particle in either compartment is 1/2. }
\label{fig:sym1}
\end{figure}

There are two fundamentally different ways to get an equal probability of occupying the left and right box of a Szilard engine. One way is to follow the usual protocol of splitting a symmetric wavefunction in two exactly equal parts, i.e. inserting a barrier in the center of a box with a particle in the ground state. Fig~\ref{fig:sym1} shows the time evolution of the eigenstates and eigenenergies when inserting a time-dependent barrier, with height $ \alpha(t) $ (dashed vertical line), in the center of the box. 1A is the initial state of the system, before the barrier has begun to be inserted. 1B is an intermediate step with $ 0<\alpha <\infty $ before the two compartments have been completely isolated from each other in 1C as $ \alpha \to \infty $. The eigenstates in 1C are split exactly in half, with a probability of 1/2 on either side.

The second way is to insert the barrier asymmetrically and non-adiabatically, in such a way that only the two first energy levels are excited; the eigenfunction of the first energy level will be large in the larger compartment and small in the smaller compartment, and vice versa for the second energy level. This method is illustrated in Fig~\ref{fig:box_illustration}. The initial state 2A, before the barrier is inserted, is identical to that of 1A. However as the barrier is increased via 2B through 2C the symmetric eigenfunctions becomes zero in the smaller compartment, while the antisymmetric becomes zero in the larger compartment. Of course it has to be this way, since when $ \alpha \to \infty $ what we have is essentially two rescaled copies of the initial state. The third energy level in 2A becomes the new first excited state of the larger compartment in 2C, while the first exited state in 2A becomes the new ground state of the smaller compartment in 2C. Exciting the second energy level of the original box still results in no excitations after the measurement, since it becomes the new ground state of the compartment.

In this article we address the two following questions: How sensitive is the non-adiabatic splitting of the wavefunction to asymmetry in barrier insertion, and how much additional entropy is produced by higher level excitations when we insert the barrier with a finite rate.

\section{Analysis}

\begin{figure}
\centering
\includegraphics[width=1\linewidth]{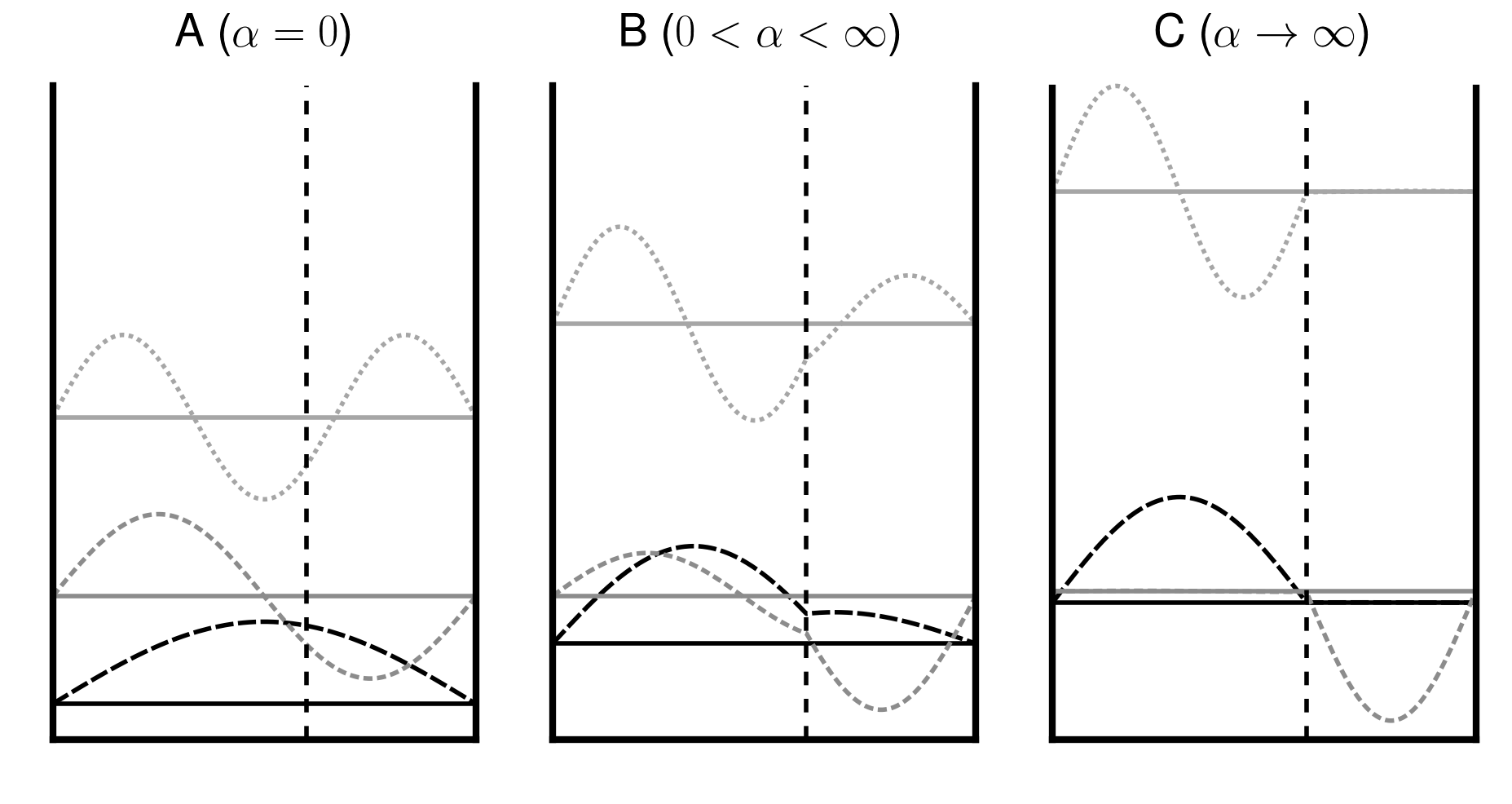}
\caption{Schematic of the three first eigenfunction and energies for an asymmetric box for three different values of the barrier height $ \alpha(t) $. A) is the initial state of the system, identical to Fig~\ref{fig:sym1}A. Only now the eigenfunction of the first excited state is non-zero at the point we insert the barrier, allowing for excitations from the ground state. From the intermediate time-step shown in B) to the final state shown in C) the eigenfunction of the ground and first excited state evolves such that it is approximately zero in the smaller and larger compartment, respectively. }
\label{fig:box_illustration}
\end{figure}

\begin{figure}
\centering
\includegraphics[width=1\linewidth]{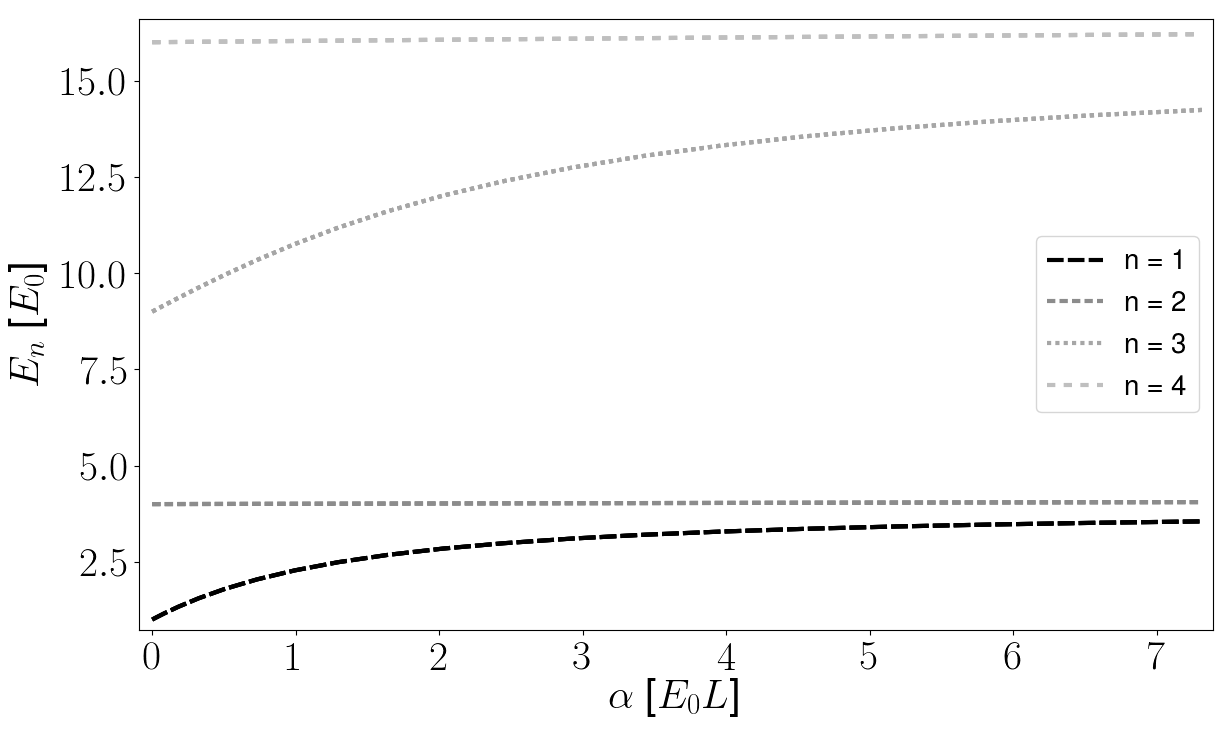}
\caption{Plot of the energy levels as a function of time. We see that the odd energy levels  approach the evens as the height of the barrier increases, and the final spacing between them decreases with the magnitude of the asymmetry.}
\label{fig:energies}
\end{figure}

The box is shown in Fig~\ref{fig:box_illustration} and is defined by the potential $ V(x) = 0$ for $ x \in [-a,b] $ and $ V(x) = \infty $ elsewhere. The barrier is a delta function with time-dependent height, $ \alpha(t) $, inserted at $ x = 0 $. We choose the barrier to be a delta function since it allows presenting the eigenstates in analytical form. A barrier with finite width was used in \cite{AmJourPhys}, while in \cite{PhysRevA.94.052124} they used a delta function barrier and obtained similar results. The width of the barrier would only affect the tunneling rate between the compartments, but the qualitative results would remain unchanged. The insertion of the barrier is described by a time-dependent Hamiltonian given by
\begin{equation}
\hat{H}(t) = -\frac{\hbar^2}{2m}\frac{\partial^2}{\partial x^2} +\alpha(t)\delta(x),
\end{equation}
where $ m $ is the mass of the particle. The instantaneous eigenfunctions, $ \ket{\psi_n(t)} $ evolve are found as the solution to the time-independent Schr\"{o}dinger equation
\begin{equation}\label{time_dep}
\hat{H}(t)\ket{\psi_n(t)}=E(t)\ket{\psi_n(t)}.
\end{equation} 
At any given time the instantaneous eigenfunctions is an orthonormal set $ \braket{\psi_n|\psi_m} = \delta_{n,m} $. Therefore the total wavefunction $ \ket{\Psi(t)} $, which is the solution of the time-dependent Schr\"{o}dinger equation
\begin{equation}
i\hbar~\partial_t \ket{\Psi(t)}=\hat{H}\ket{\Psi(t)},
\end{equation}
can be expressed as a linear combination of them
\begin{equation}\label{total_wave}
\ket{\Psi(t)}=\sum_{n}c_n(t)\ket{\psi_n(t)}e^{i\theta_n(t)},\quad \theta_n = -\frac{1}{\hbar}\int_{0}^{t}E_n(t')dt'.
\end{equation}
Here $ c_n(t) $ is a set of complex constants satisfying $ \sum_{n}^{\infty} |c_n(t)|^2 =1$ . As shown the appendix \ref{wave_function}, the system of coupled differential equations giving the time-evolution of the coefficients $ \{c_n\} $ is
\begin{equation}\label{diff_eq_text}
\dot{c}_n(t) = - \sum_{m\neq n} c_m(t)\frac{\braket{\psi_n(t)|\partial_t\hat{H}|\psi_m(t)}}{E_m-E_n}e^{i(\theta_m-\theta_n)}.
\end{equation}
We first need to find the instantaneous solutions $ \ket{\psi_n(t)} $ for the asymmetric barrier problem, and the details of these calculations are given in appendix \ref{asymmetric_wall}. After finding the instantaneous solutions we numerically solve Eq.~(\ref{diff_eq_text}) to find the time-evolution of $ \ket{\Psi(t)} $.

\section{Results}
Let us now see to what extent it is possible to make the probability to find the particle in either compartment equal (or as close to equal as possible), while limiting excitations to higher energy states.\\

We set the total length of the box equal to $ L= a + b = 1 $, and define $ a = 1/2 + \epsilon $, where $ \epsilon $ is the asymmetry parameter that determines how much larger the compartment on the left side of the barrier is than the one on the right side. We also set $ \hbar = m = 1 $. The initial state is chosen to be the ground state, that is $ c_1(0) = 1 $ and $ c_n(0) = 0$ for $ n>1 $. We found that including the six first eigenstates was sufficient capture all the excitations for the insertion rates we explored. We set the maximum height of the barrier at the end of the protocol ($ t=\tau $) to $ \alpha(\tau) = 400 ~E_0 $, where $ E_0 $ is the ground state of the box of $ L = 1 $ without a barrier. This value was chosen to make sure that the coefficients $\{ c_n(\tau)\} $ have converged to a constant values.\\

For the protocol we chose $ \alpha(t) = At^2 $, where $ A $ is some constant that determines the rate of insertion. We also tried a linear protocol, but found that in order to limit higher order excitation the rate of insertion had to start small and steadily increase as a function of time.
\begin{figure}
\centering
\includegraphics[width=1\linewidth]{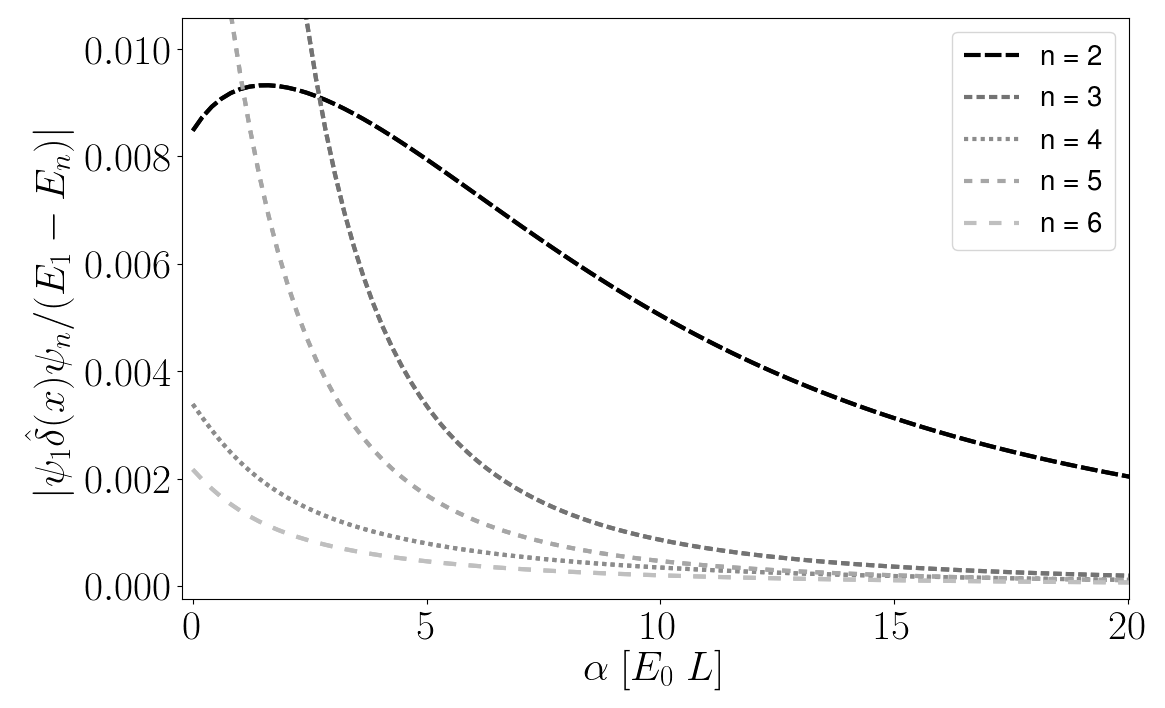}
\caption{In this plot we show how the ratio $ \frac{\braket{\psi_1(t)|\hat{\delta}(x)|\psi_m(t)}}{E_m-E_1} $ depends on the height of the barrier $ \alpha $. Its magnitude gives us an indication of the coupling between ground state and the higher excited states. We see that the coupling between the ground state and the first excited state remains substantial for high values of $ \alpha $, while all the others decay quickly. This indicates that we can induce transitions between those two levels without exiting higher states when $ \alpha $ is large. This plot was obtained with $ \epsilon $ = 0.1}
\label{fig:overlap_factor}
\end{figure}
The reason for this can be understood by studying the coupling between the $ \{c_n(t)\} $ in Eq.(\ref{diff_eq_text}) at a given time $t$
\begin{equation}\label{coeff}
\frac{\braket{\psi_n(t)|\partial_t\hat{H}|\psi_m(t)}}{E_m-E_n} =\dot{\alpha}(t) \frac{\braket{\psi_n(t)|\hat{\delta}(x)|\psi_m(t)}}{E_m-E_n}.
\end{equation}
When we insert the barrier, the probability to find the particle at the insertion point decreases in proportion to the height of the barrier. Therefore numerator, $ \braket{\psi_n(t)|\hat{\delta}(x)|\psi_m(t)} $, which measures overlap between the eigenstates at the insertion point, will be largest in the beginning and decrease towards zero as the barrier height is increased. This prevents transitions for high barriers. The denominator is the energy difference between the eigenstates, $ E_m-E_n $, and its dependence on the barrier height is shown in Fig~\ref{fig:energies}. The energy difference between the ground state and the first exited state is largest in the beginning and asymptotically approaches a final small value that increases with the asymmetry between the compartments. This makes transition between these more likely as the barrier height increases.

In Fig.~\ref{fig:overlap_factor} we plot the ratio $  \braket{\psi_1(t)|\hat{\delta}(x)|\psi_m(t)}/(E_m-E_1) $ and interpret its magnitude as an indication of the coupling strength between the ground state and the $ m $th eigenstate. As argued in the previous paragraph we see that indeed the ground state coupling to the first excited state dominates over its coupling to other eigenstates once the barrier has reached a certain height ($  \simeq 4 ~E_0 $ in this example, where $ \epsilon= 0.1 $). As seen in Eq.~(\ref{coeff}), we can control the coupling strength via $ \dot{\alpha}(t) $. By choosing a $ \dot{\alpha}(t) $ that is small in the beginning and large towards the end of the protocol, we suppress early transitions between the levels when $ \braket{\psi_n(t)|\hat{\delta}(x)|\psi_m(t)} / (E_m-E_n) $ is large. Since the energy difference between the ground state and the first excited state becomes much smaller than than the difference between the ground state and any of the higher states, we can induce transitions between the them, even when the wavefunction overlap is very small, if we choose a $ \dot{\alpha}(t) $ that is suitably large.\\

\begin{figure}[t]
	\centering
	\includegraphics[width=\linewidth]{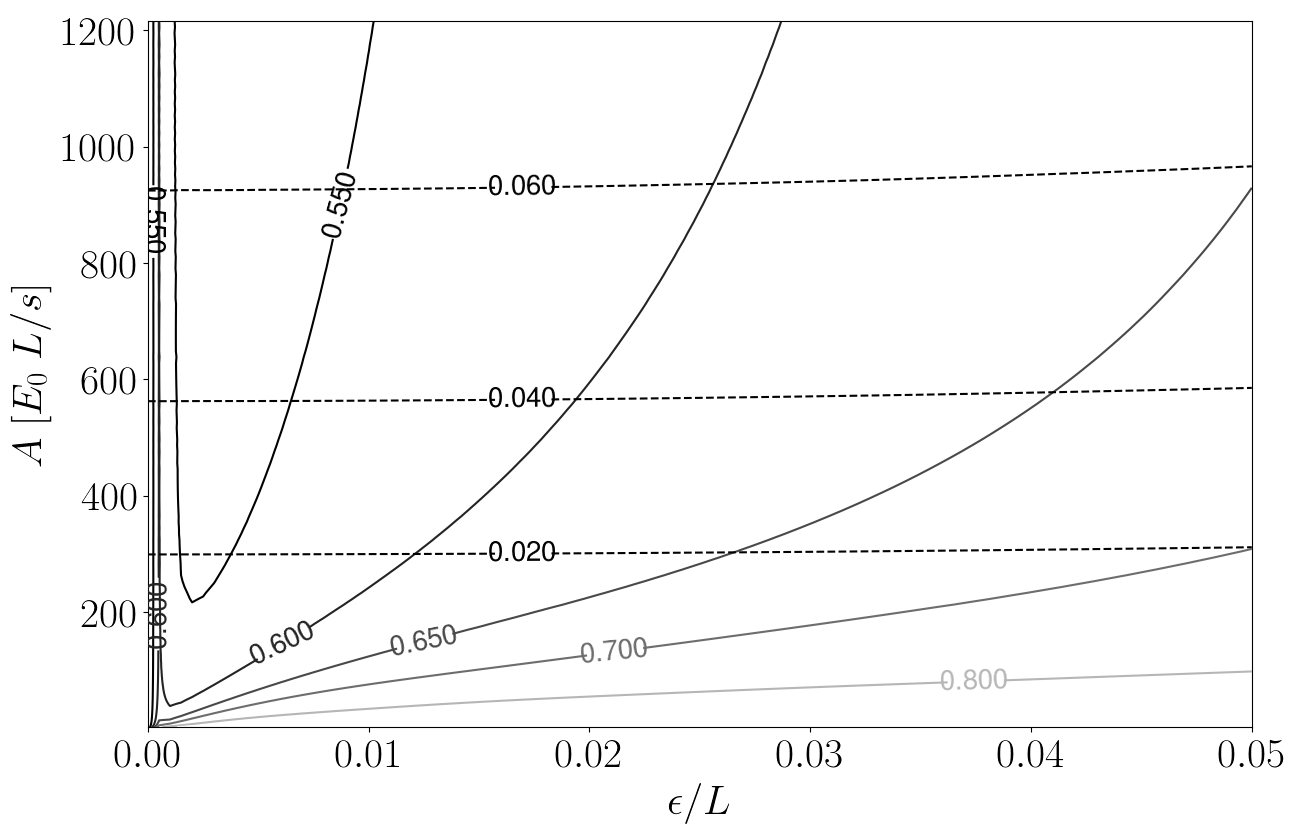}
	\caption{Contour plot of the probability to find the particle in the largest compartment (solid lines), as a function of the barrier insertion rate constant $ A $ and the asymmetry parameter $ \epsilon $. The probability to excite levels higher than the second is shown in the dashed lines.}
	\label{fig:P_left_contour}
\end{figure}

In Fig~\ref{fig:P_left_contour} we show a contour plot of the probability to find the particle in the bigger compartment (solid lines) at the end of the protocol as function of the asymmetry parameter $ \epsilon/L $ and the insertion rate parameter $ A $. We see that even for asymmetries of the order of $ \epsilon \sim 0.01  $ the probability to find the particle in the bigger compartment is quite large.
Although increasing the barrier faster makes the probabilities to find the particle in either side more equal it also incurs a penalty; the faster you increase the barrier the more likely it is that you excite higher order states in the energy spectrum. Higher order excitations increases the entropy of the system, since the internal states of the Szilard engine is assumed to be either the ground state (bigger compartment) or the first excited state (smaller compartment).

\section{Summary and Discussion}
When designing a Szilard engine one wants the probabilities to find the particle in either compartment after barrier insertion to be equal. Experimentally it might be difficult to design a perfectly symmetric double well potential. We have answered the question of how sensitive the probability distribution of the divided single-particle box is to asymmetry between the compartment size. It turns out that even for very small differences between the width of the compartments, the resulting probability distribution is heavily skewed towards the larger compartment. The faster one increases the barrier height, the more even the final distribution becomes. However, this rapid increase also leads to higher order excitations in the box, which results in unwanted entropy production. We have used a simple function that is quadratic in time for the height of the barrier, and similar results to that which we have shown is valid for linear and other similar functions that have monotonously increasing derivatives. We point out the fact that excitations to the second level are special in the sense that after the barrier height becomes high enough to stop tunneling between the two compartments, and a measurement to determine which compartment the particle is found is performed, the system is still in the ground state for the relevant compartment. The question remains whether a protocol for $ \alpha(t) $ can be constructed such that the interference of the eigenfunctions results in an equal final distribution between the left and right side, even if there is asymmetry between the compartment sides.

\section*{Acknowledgements}
We would like to thank Y. M. Galperin for valuable discussions and comments on the manuscript.

\appendix
\section{Wave function for time-dependent Hamiltonian}\label{wave_function}
In this section we follow \cite{griffiths_schroeter_2018} (section 10.1.2) and write the total wavefunction $ \ket{\Psi (t)} $ as a linear combination of the instantaneous eigenstates $ \ket{\psi(t)_n} $ and derive the coupled differential equation for the coefficients. When the Hamiltonian changes with time, the eigenfunctions and eigenvalues are also time-dependent,
\begin{equation}\label{time_dep}
\hat{H}(t)\ket{\psi_n(t)}=E_n(t)\ket{\psi_n(t)}.
\end{equation}
The eigenfunctions at any given time is an orthonormal set,  $ \braket{\psi_n(t)|\psi_m(t)}=\delta_{n,m} $, and the total wavefunction which can be found as the solution of the time-dependent Schr\"{o}dinger equation
\begin{equation}
i\hbar~\partial_t \ket{\Psi(t)}=\hat{H}\ket{\Psi(t)},
\end{equation}
can be expressed as a linear combination of them:
\begin{equation}\label{total_wave}
\ket{\Psi(t)}=\sum_{n}c_n(t)\ket{\psi_n(t)}e^{i\theta_n(t)},
\end{equation}
where
\begin{equation}
\theta_n = -\frac{1}{\hbar}\int_{0}^{t}E_n(t')dt'.
\end{equation}
Inserting this linear combination into the time-dependent Schr\"{o}dinger equation gives us
\begin{eqnarray}
&i\hbar\sum_n\left[\dot{c}_n\ket{\psi_n} + c_n\ket{\dot{\psi}_n} + ic_n\ket{\psi_n}\dot{\theta}_n\right]e^{i\theta_n} \\
&= \sum_nc_n \hat{H}\ket{\psi_n}e^{i\theta_n}.
\end{eqnarray}
Now since $ \dot{\theta}_n = -E_n/\hbar $ and $ \hat{H}\ket{\psi_n} = E_n\ket{\psi_n} $, the right hand side exactly cancels the last term on the left hand side and we are left with
\begin{equation}
\sum_n\left[\dot{c}_n\ket{\psi_n} + c_n\ket{\dot{\psi}_n}\right]e^{i\theta_n} = 0.
\end{equation}
We now take the inner product with the eigenfunction $ \psi_m $, and since the eigenfunctions constitute an orthonormal set at any given time $ t $, we obtain a set of $ N $ coupled differential equations for the $ N $ coefficients $ c_n, ~n\in[1,N] $.
\begin{eqnarray}
\sum_{n}\left[ \dot{c}_n\delta_{m,n}+c_n\braket{\psi_m|\dot{\psi}_n}\right] e^{i\theta} = 0\\
\label{diff_eq}
\dot{c}_m(t)=-\sum_{n}c_n\braket{\psi_m|\dot{\psi}_n}e^{i(\theta_n-\theta_m)}.
\end{eqnarray}
We can rewrite this equation by taking the time derivative of Eq.~\ref{time_dep} and then the inner product with $ \psi_m $ to obtain
\begin{equation}
\braket{\psi_m|\dot{\hat{H}}|\psi_n}+E_m\braket{\psi_m|\dot{\psi}_n}= \dot{E}\delta_{m,n}+E_n\braket{\psi_m|\dot{\psi}_n},
\end{equation}
which shows us that the inner product $ \braket{\psi_m|\dot{\psi}_n} $ can be written as
\begin{equation}
\braket{\psi_m|\dot{\psi}_n} = \frac{\braket{\psi_m|\dot{\hat{H}}|\psi_n}}{E_n-E_m},
\end{equation}
as long as the system is non-degenerate and $ n\neq m $. Putting this result into Eq.~\ref{diff_eq} we get
\begin{equation}
\dot{c}_m = - c_m \braket{\psi_m|\dot{\psi}_m} - \sum_{n\neq m} c_n\frac{\braket{\psi_m|\dot{\hat{H}}|\psi_n}}{E_n-E_m}e^{i(\theta_n-\theta_m)}.
\end{equation}
This form of the differential equation is particularly well suited to our problem. Firstly, the Hamiltonian contains a delta-function at $ x = 0 $, so the integral $ \braket{\psi_m|\dot{\hat{H}}|\psi_n} $ is simply given by (using the eigenfunctions from appendix \ref{asymmetric_wall})
\begin{equation}
\braket{\psi_m|\dot{\hat{H}}|\psi_n} = \dot{\alpha}A_nA_m\sin(k_na)\sin(k_ma).
\end{equation}
In addition, the term $ \braket{\psi_m|\dot{\psi}_m} $ is always zero. This is because the instantaneous eigenfunctions $ \ket{\psi_m} $ are orthonormal ($ \braket{\psi_m|\psi_m} = 1 $) and real.
\begin{equation}
\frac{\partial}{\partial t}\braket{\psi_m|\psi_m} = \braket{\dot{\psi}_m|\psi_m} + \braket{\psi_m|\dot{\psi_m}}= 0.
\end{equation}
Since $ \braket{\psi_m|\dot{\psi}_m} = \braket{\dot{\psi}_m|\psi_m}^* $ we get
\begin{equation}
\braket{\psi_m|\dot{\psi}_m} = - \braket{\psi_m|\dot{\psi}_m}^*\to \mathtt{Re}[\braket{\psi_m|\dot{\psi}_m}] = 0.
\end{equation}
Therefore the coupled differential equations we need to solve become 
\begin{equation}\label{diff_eq_final}
\dot{c}_m = - \sum_{n\neq m} c_n\frac{\braket{\psi_m|\dot{\hat{H}}|\psi_n}}{E_n-E_m}e^{i(\theta_n-\theta_m)}.
\end{equation}

\section{Asymmetric barrier}\label{asymmetric_wall}
We can find the stationary states from the time-independent Schr\"{o}dinger equation and they have the form
\begin{equation}
\psi(x)=\left\{
\begin{array}{ll}
A\sin\left[k(x+a)\right],\qquad x\in[-a,0]\\
B\sin\left[k(x-b)\right],\qquad x\in[0,b]
\end{array}
\right.
\end{equation}
where $ k = \sqrt{2mE}/\hbar $. At $ x = 0 $ the wavefunction is continuous while its derivate has a discontinuity. These two conditions are 
\begin{eqnarray}
\lim_{\epsilon\to 0}\left[\psi(0-\epsilon)-\psi(0+\epsilon)\right] &=& 0,\\ \lim_{\epsilon\to 0}\left[\dot{\psi}(0+\epsilon)-\dot{\psi}(0-\epsilon)\right] &=& \frac{2m\alpha}{\hbar^2}\psi(0),
\end{eqnarray}
and for our system they result in 
\begin{eqnarray}\label{continuity}
A\sin(ka)&=& -B\sin(kb),\\
B\cos(kb)-A\cos(ka)&=&\frac{2m\alpha}{k\hbar^2}A\sin(ka).
\end{eqnarray}
Combining these equations gives us another one, that we can solve numerically to find the wavevectors $ k $ for a given $ a,b $ and $ \alpha $. 
\begin{equation}
\sin(k(a+b))=-\frac{2m\alpha}{k\hbar^2}~\sin(ka)~\sin(kb).
\end{equation}
The solutions to this equation defines a discrete set of allowed values for the wavevector $ k\to k_n, ~n = 1,2,\dots $ which determines the energy spectrum of the system via 
\begin{equation}
E_n = \frac{\hbar}{2m}k_n^2.
\end{equation}
The wavefunction has to be normalized on the domain of $x$,
\begin{equation}
\int_{-a}^{0}A_n^2\sin^2\left[k_n(x+a)\right]+\int_{0}^{b}B_n^2\sin^2\left[k_n(x-b)\right] = 1
\end{equation}
which combined with Eq.~\ref{continuity} gives us the normalization constants $ A_n $
\begin{equation}
A_n^2 = \left[\frac{a}{2}-\frac{\sin(2k_na)}{4k_n} + \frac{\sin^2(k_na)}{\sin^2(k_nb)}\left(\frac{b}{2}-\frac{\sin(2k_nb)}{4k_n}\right)\right]^{-1}.
\end{equation}
The $ B_n $'s can be found via Eq.~(\ref{continuity}).

\end{document}